# Combining Radiomics and Machine Learning Approaches for Objective ASD Diagnosis: Verifying White Matter Associations with ASD

Junlin Song[1#], Yuzhuo Chen[1#], Yuan Yao[3#], Zetong Chen[1], Renhao Guo[1], Lida Yang[1], Xinyi Sui[1], Qihang Wang[1], Xijiao Li[1], Aihua Cao[2*], Wei Li[1*]

[1] School of Control Science and Engineering, Shandong University, Jinan City, Shandong Province, 250061, China

[2] Department of Pediatric, Qilu hospital of Shandong University, Jinan City, Shandong Province, 250012, China

[3] Department of Radiology, Qilu Hospital of Shandong University, Jinan City, Shandong Province, 250012, China

#These authors contributed to the work equally and should be regarded as co-first authors.
**Corresponding authors**: Wei Li, Aihua Cao

Wei Li
School of Control Science and Engineering, Shandong University
Address: Qianfoshan Campus, Shandong University, 17923 Jingshi Road, Jinan City, Shandong Province, 250061, China
Telephone number: +86-15153150160
E-mail: cindy@sdu.edu.cn

Aihua Cao
Cheeloo College of Medicine, Shandong University, No. 44, Wenhua West Road, Lixia District, Jinan City, Shandong Province, 250012, China
Address: Qilu hospital of Shandong University, Jinan City, Shandong Province, 250012, China
Shandong Province, 250012, China
Telephone number: +86-018560086317
E-mail: xinercah@163.com

# Combining Radiomics and Machine Learning Approaches for Objective ASD Diagnosis: Verifying White Matter Associations with ASD

**Abstract**

Autism Spectrum Disorder is a condition characterized by a typical brain development leading to impairments in social skills, communication abilities, repetitive behaviors, and sensory processing. There have been many studies combining brain MRI images with machine learning algorithms to achieve objective diagnosis of autism, but the correlation between white matter and autism has not been fully utilized. To address this gap, we develop a computer-aided diagnostic model focusing on white matter regions in brain MRI by employing radiomics and machine learning methods. This study introduced a MultiUNet model for segmenting white matter, leveraging the UNet architecture and utilizing manually segmented MRI images as the training data. Subsequently, we extracted white matter features using the Pyradiomics toolkit and applied different machine learning models such as Support Vector Machine, Random Forest, Logistic Regression, and K-Nearest Neighbors to predict autism. The prediction sets all exceeded 80% accuracy. Additionally, we employed Convolutional Neural Network to analyze segmented white matter images, achieving a prediction accuracy of 86.84%. Notably, Support Vector Machine demonstrated the highest prediction accuracy at 89.47%. These findings not only underscore the efficacy of the models but also establish a link between white matter abnormalities and autism. Our study contributes to a comprehensive evaluation of various diagnostic models for autism and introduces a computer-aided diagnostic algorithm for early and objective autism diagnosis based on MRI white matter regions.

**Keywords**

ASD; MRI; Machine Learning; radiomics; White Matter

## 1.Introduction

Autism Spectrum Disorder (ASD) is a developmental condition characterized by deficits in social skills and restricted and repetitive behaviors, interests, or activity patterns that may persist throughout an individual's life[1]. The prevalence of ASD in the United States is currently estimated at 1 out of 36 individuals[2]. The exact cause of ASD remains unknown, with the majority of cases believed to result from a complex interplay of genetic and environmental factors. The intricate nature of ASD, coupled with limited understanding of its underlying causes and biochemical abnormalities, presents challenges in both diagnosis and treatment, making it a global concern[3][4][5].

While there is no specific pharmaceutical cure for ASD, early diagnosis plays a crucial role in improving patients' quality of life through timely and appropriate interventions. Therefore, there is an urgent need to develop accurate methods for early ASD diagnosis[6][7][8]. In clinical practice, physicians typically rely on the Diagnostic and Statistical Manual of Mental Disorders 5 (DSM-5) criteria to identify social interaction deficits, restricted interests, stereotypical behaviors, and sensory interests as diagnostic indicators. However, the autism scale is qualitative which do not fully reflect the complexity of ASD. It may not be diagnosed until patients present with social deficits and behavioral patterns, which makes it easy to miss out on the optimal time for therapeutic interventions for ASD[9][10].

Recognizing the limitations of traditional diagnostic approaches, there is a growing emphasis on utilizing neuroimaging techniques for ASD diagnosis. Magnetic Resonance Imaging (MRI) has emerged as a valuable, non-invasive tool for studying the human brain due to its high spatial resolution and safety[11][12]. Studies utilizing structural MRI have identified brain structure alterations in ASD patients, such as increased frontal and temporal lobe volumes, enlarged intracranial volumes, and differences in cortical shape compared to neurotypical individuals[13][14][15]. These findings underscore the potential of brain MRI analysis in identifying biomarkers of autism and laying the groundwork for the development of computerized diagnostic platforms based on neuroimaging data.

The rise of artificial intelligence has led to a growing interest in its application within the medical field, including utilizing MRI images to predict Autism Spectrum Disorder (ASD).Taban et al.[16] develop an auto-ASD network, which fed the features extracted by MLP into an SVM classifier, and achieved a classification accuracy of over 70% for 4 fMRI datasets, with the highest accuracy reaching 80%.Ahmed et al.[17] established a model consisting of a restricted Boltzmann machine (RBM) and a support vector machine (SVM), which were used to extract features and classify ASD subjects, respectively. The features extracted by RBM achieved an accuracy of 83% in the classifier. Sara[18] extracted 221 brain morphological features and entered them into a random forest (RF) model for training to achieve the diagnosis of autism. After data coordination, the highest AUC of the RF classifier reached 75%. Ismail[19] used the cortical morphology network derived from T1-weighted MRI to diagnose ASD and achieved an accuracy of 70%, a sensitivity of 72.5%, and a specificity of 67.5%. Matthew.et al.[20] used the largest multi-source functional MRI (fMRI) connectome dataset (consisting of 43,858 data points) to train a convolutional neural network (CNN) and achieved an overall AUROC of 67.74%, 76.80%, and 92.22% for ASD, normal, gender, and task and rest.

Although there have been considerable studies on combining machine learning and MRI image training classifiers to achieve autism diagnosis, most of the methods use the whole brain MRI image as input without separating specific brain structures to train the classifier. Given the absence of research utilizing white matter in brain MRI for training classification networks in autism diagnosis, and the differences in white matter structure between ASD patients and normal people have been proposed in some studies[21][22][23], this study segmented the white matter portion of the brain MRI images. Features were then extracted using radiomics methodology and inputted into respective models for training, with the aim of assessing the influence of white matter abnormalities on autism.

In this study, we employed several models trained on brain white matter features for comprehensive comparison. These models cover widely used algorithms in the field of machine learning such as Support Vector Machines(SVM), Convolutional Neural Network (CNN), Random Forest (RF), K-Nearest Neighbor(KNN), Logistic Regression (LG).Through the training and evaluation of these diverse machine learning models, a more thorough understanding of their efficacy in ASD classification tasks was achieved. This approach facilitated the identification of the strengths and limitations of different algorithms in processing brain white matter features for ASD classification, aiding in the selection of an optimal model for ASD diagnosis. Furthermore, by employing multiple models in this context, the study was able to discern and compare their capacity to capture the relationship between white matter abnormalities and ASD. This methodology served to validate the significance of white matter lesions in autism and offer deeper insights into the underlying pathophysiological mechanisms of ASD.

## 2.Material and methods

### 2.1.Data Acquisition

#### 2.1.1.Participants

The study involved children aged 2-5 years, consisting of both healthy individuals and those with ASD. There were a total of 62 healthy participants and 85 children with ASD, all recruited from Qilu Hospital of Shandong University. The children with ASD met the diagnostic criteria outlined in the DSM-5 and Childhood Autism Rating Scale (CARS). Exclusion criteria for the study included individuals with a history of seizures, head trauma, genetic or neurological diseases, major illnesses, general anesthesia exposure, etc. Parental consent was obtained for all participants, and the research procedures were approved by the Institutional Review Board of Qilu Hospital, Shandong University.

#### 2.1.2.Magnetic resonance imaging protocol

We acquired magnetic resonance images using a standard orthogonal head coil on the Magnetom Verio 3.0 device (manufactured by Siemens, Germany) at Qilu Hospital, Shandong University. For the purpose of this study, we acquired high-resolution images for thick-slice MRI based on T1-weighted (T1-MRI) for surface-based analysis: TR=1900 milliseconds, TE=8.5 milliseconds; flip angle=150°; FOV=220 millimeters; slice thickness=6.0 millimeters; ACQ yoxel MPS=0.85*1.13*6.00 millimeters; matrix= 256×256; number of slices=18; number of excitations=1.0. The scanned images were all reviewed for quality by experienced radiologists.

### 2.2.Data Preprocessing

The outcomes of each preprocessing stage are depicted in Fig.1. Initially, noise removal filtering was conducted, as illustrated in A, to eliminate artifacts interference in the image. Subsequently, image registration was performed, as shown in B, to mitigate the impact of variations in MRI image shape and size on subsequent model training. Gray scale normalization and histogram equalization were also carried out to enhance the image visual quality and contrast, aiding in delineating white matter and training the segmentation model. Ultimately, segmentation was executed in ITK-SANP, as displayed in D, to generate samples for training the segmentation network.

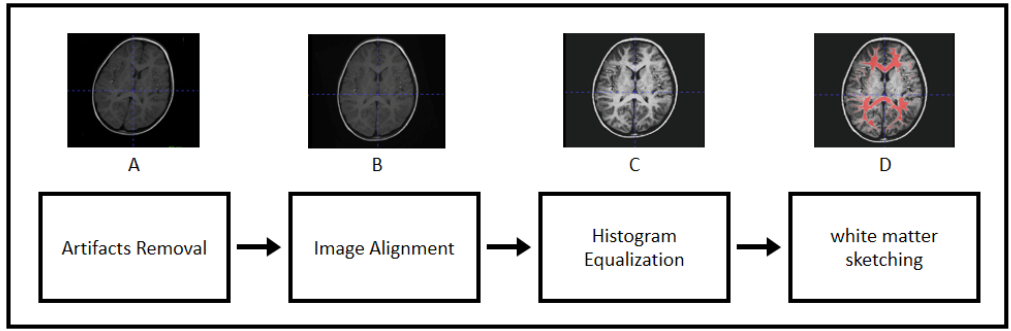

Fig.1. Preprocessing flow-chart

### 2.2.1. Artifacts Removal

Artifacts in MRI refer to abnormal, distorted, or false signal characteristics that appear in MRI images and usually do not represent the true anatomical structure. Movement or breathing during scanning can cause blurring or distortion of the image, resulting in motion artifacts. Uneven magnetic fields can also cause local signal intensity changes or image distortion, resulting in artifacts. These artifacts can seriously degrade image quality and interfere with the correct identification and analysis of image anatomy. Therefore, noise reduction filters was utilized to eliminate the interference of artifacts on the image.

### 2.2.2. Image Alignment

Image alignment is the process of aligning similar features in two or more images so that they correspond in space. The MRI head scan takes between ten and thirty minutes, depending on the procedure. During the scan, subjects may inadvertently sway their head, causing some MRI images to have a large tilt. In addition, each MRI image varies in shape and size. To effectively compare different locations in image space, Image alignment preprocessing were performed. Using the brain image spatial alignment function of the CAT12 software, all MRI images were registered to a uniform standard space.

### 2.2.3. Histogram Equalization

Histogram Equalization (HE)[24] is a method in digital image processing that enhances image contrast by redistributing the grayscale values to achieve a more balanced distribution across the brightness range. This technique aims to improve the visual quality and contrast of the image by highlighting details, making it easier to observe and analyze structures and anomalies. In our study, applying histogram equalization has proven beneficial in enhancing the contrast of brain MRI images, particularly in making the boundaries between white matter and other tissues more distinct. This improvement aids the segmentation network in accurately identifying white matter regions.

**2.2.4. Grayscale transformation**

After histogram equalization, artifacts may reappear due to the migration of high-intensity pixels to low-intensity pixels in the histogram. This manifests itself as white artifacts in the low-intensity areas of the image. Based on this, the single-threshold grayscale transformation was utilized to set the grayscale values below the set threshold to 0 and map the grayscale values above the set threshold to the range 0-255.

Grayscale transformation formula:

$$y = \begin{cases} 0 & , x < \text{threshold} \\ \frac{x - \text{threshold}}{\max - \text{threshold}} \times 255 & , x \geq \text{threshold} \end{cases} \quad (2)$$

Where y is the transformed grayscale, x is the untransformed grayscale, threshold is the cutoff threshold, and max is the maximum grayscale in the untransformed image. After several attempts, we found the optimal threshold to be 100. The grayscale transformation curve is shown in Fig.2.

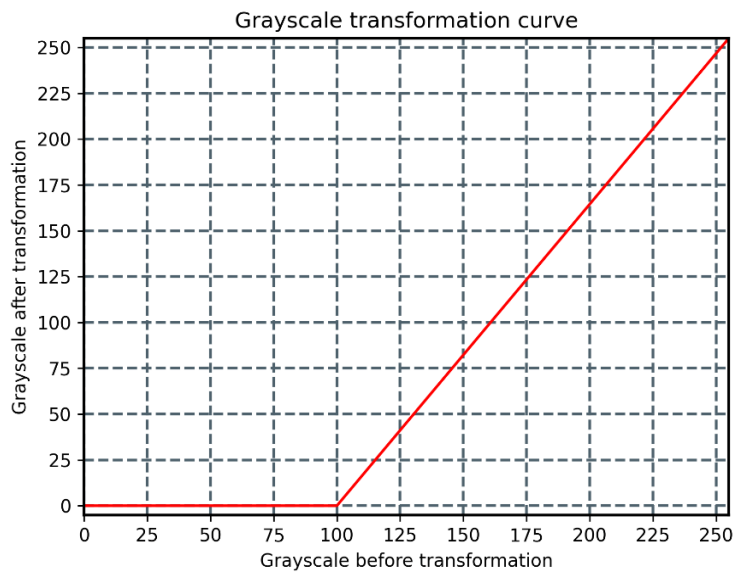

Fig.2. Grayscale transformation Curve

**2.3. Segmentation model establishment**

**2.3.1. UNet**

To explore the impact of white matter lesions on ASD, this study utilized MRI image data manually segmented by physicians as ground truth to train a segmentation model for automatic white matter segmentation. Currently, medical image segmentation methods mainly include traditional methods and deep learning methods. Traditional medical image segmentation methods mainly rely on features such as grayscale and texture of medical image lesion areas through thresholding, edge detection, region growing, active contour, fuzzy clustering, and wavelet transform[25][26][27]. However, conventional segmentation algorithms have poor performance in recognizing the spatial distribution characteristics of image pixel grayscale values and image edge features, which makes segmentation algorithms sensitive to noise and not suitable for our study.

Deep learning based methods can automatically discover image feature information through different network structures, combine shallow feature information with deep feature information, and achieve good segmentation results, and are developing in the end-to-end direction[28][29]. UNet[30]is a classic deep learning image segmentation network that can solve pixel-level semantic segmentation tasks in medical image segmentation, such as cell segmentation and organ segmentation.The structure of the UNet network is shown in Fig.3.It consists of a contracting path (left) and an expanding path (right).

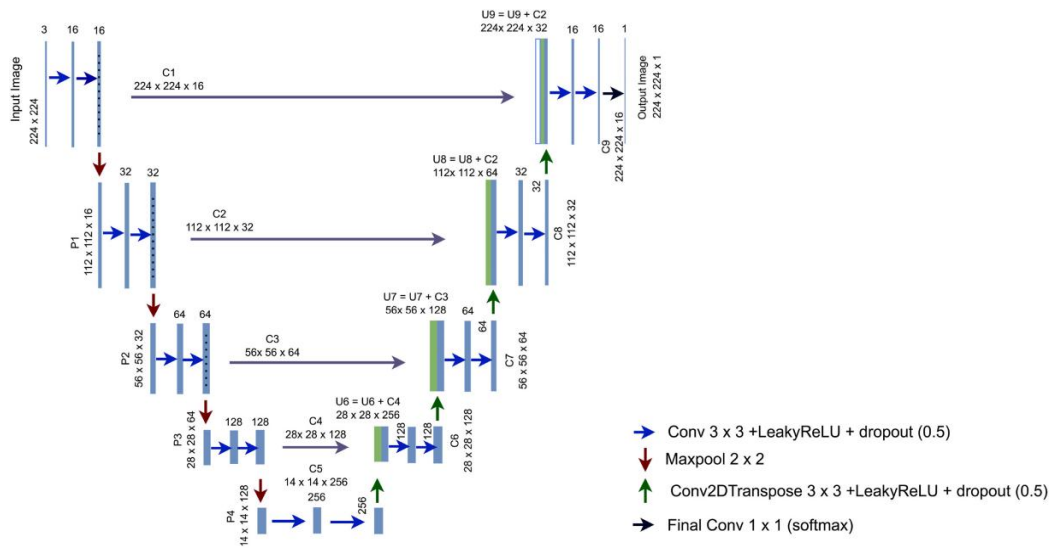

Fig.3.UNet Structure

We utilized manually segmenteds mask as ground truth and preprocessed images as the input data. The UNet model is trained with the training set examples, and then the prediction set examples are fed into the trained UNet to generate the predicted mask.

### 2.3.2.MultiUNet

Based on UNet, this study proposed a MultiUNet for white matter segmentation task to achieve more accurate and reliable segmentation of white matter regions in thick-slice MRI images. The network structure is shown in Fig.4.

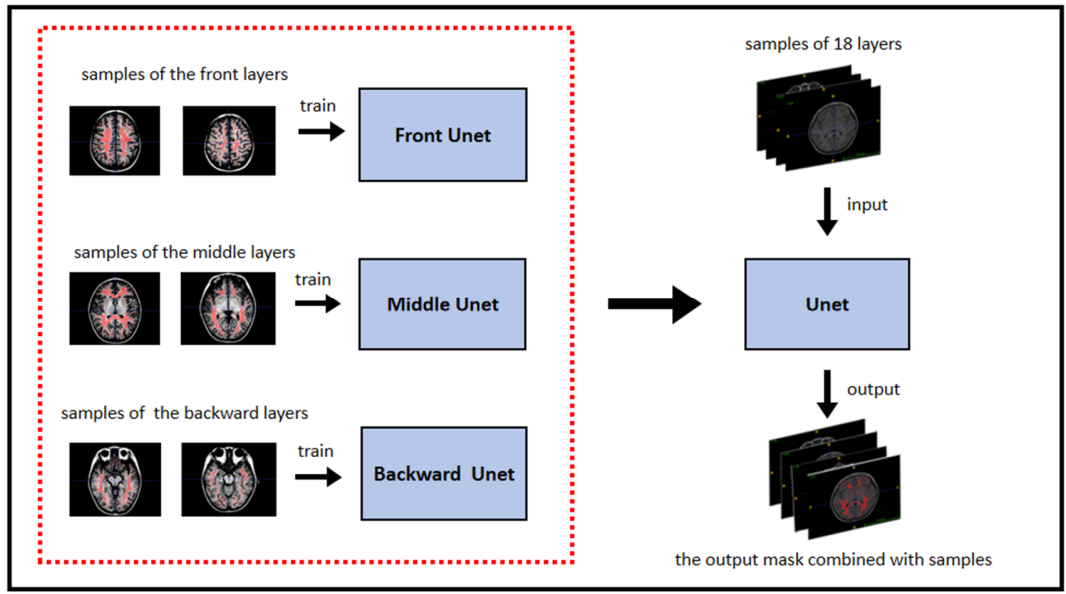

Fig.4.Our MultiUNet Structure

Due to the substantial gaps in sampling intervals and the distinct morphological variations among layers in thick-slice MRI images, utilizing all 18 layers for training the segmentation network could result in notable segmentation inaccuracies. To tackle this issue, we created the MultiUNet model, comprising three UNet models operating in parallel, with each responsible for segmenting six layers from the front, middle, and back sections of the thick-slice MRI images. This unique model structure enabled the efficient extraction of features within each section of the MRI images.

### 2.4. Feature Extraction

The Pyradiomics toolkit is utilized to extract radiomics features from medical images. In this study, Pyradiomics was employed to extract white matter features from brain MRI images to analyze the depth information within the white matter. The extracted features from the training set were used to train a classification model, while features from the test set were used for model test. The radiomic features extracted included various types such as First Order Features, Shape Features, Gray Level Co-occurrence Matrix, Gray Level Size Zone Matrix, Gray Level Run Length Matrix, and Gray Level Dependence Matrix. These features captured grayscale and shape information from the images, aiding in distinguishing between ASD and normal samples. After feature extraction, the study assessed the impact of different features on ASD recognition and selected a 108-dimensional feature set for classification.

### 2.5. Classification model establishment

To comprehensively study how various models perform in classifying ASD using white matter from brain MRI scans and to create an effective system for diagnosing ASD, we utilized a range of common machine learning algorithms including SVM, CNN, RF, LR, and KNN to thoroughly evaluate their effectiveness.

**2.5.1.SVM**

Support Vector Machine (SVM)[31] is a commonly used supervised learning algorithm that achieves the classification task by constructing an optimal hyperplane. SVM is well-suited for our research as it has a good performance in dealing with high-dimensional small-sample data and feature selection, which overcomes the demands of large samples and effectively solves the the issues of the curse of dimensionality and local minimum. Currently, the method of combination of SVM with brain MRI has been applied in various diseases, including autism spectrum disorder (ASD) [32], Alzheimer's disease (AD) [33], Parkinson's disease (PD) [34], and Depression[35], etc. SVM is capable of classifying not only linearly, but also non-linearly using kernel functions. To avoid the risk of overfitting the data, a soft-edge linear kernel SVM is used in this study due to the relatively limited number of subjects considered.

Support Vectors after introducing slack variables:

$$\omega \cdot x_1 + b \geq 1 - \xi_i, iy_i = +1 \qquad (4)$$
$$\omega \cdot x_2 + b \geq 1 - \xi_i, iy_i = -1$$

**2.5.2.RF**

Random Forests (RF)[36] is an ensemble learning method that integrates multiple decision trees. Each decision tree is constructed by randomly sampling data and selecting features for training.When constructing each tree, we utilized a different bootstrap sample(random and replacement extraction) for the training set. For each tree, approximately one third of the training instances are not used in its generation,and they are called OOB(out-of-bag sample) for this tree. Serving as an unbiased estimate of the generalisation error of the RF, the OOB misclassification rate approximates to k-fold cross-validation that requires a lot of computation. Consequently, this method allows for the selection of the best number of features by comparing OOB misclassification rate.

**2.5.3.KNN**

K-Nearest Neighbors (KNN) [37] is an instance-based classification algorithm that determines the classification of a new sample by calculating the distance between samples. The key idea is to assume that similar samples in feature space have similar have similar class labels, and there for the new sample is assigned to the majority class to which their closest K samples belong. The advantage of the KNN algorithm in ASD classification lies in its non-parametric nature, which can accommodate the variety of irregular feature distributions and decision boundary patterns.

**2.5.4.LR**

Logistic Regression (LR) [38] is a classical linear classification algorithm that first fits a decision boundary (not limited to linear, but also polynomial) and then establishes a probabilistic link between this boundary and the classification, thereby obtaining the probability of the binary classification. It not only predicts the category but also obtains the probability of that prediction, which is useful for some tasks that utilize probability to assistant decision-making. LR can be used in ASD classification to construct simple and highly interpretable models. The coefficients can be used to evaluate the importance of different features for ASD spectrum disorders, providing interpretation and understanding of these features.

### 2.5.5. CNN

Convolutional Neural Network (CNN)[39] is a deep learning algorithm particularly suitable for image recognition and processing tasks. It can capture complex spatial patterns in images in ASD classification and is highly effective in preserving spatial contextual information. CNN is a type of multilayer neural network with a unique weight sharing structure that greatly reduces the complexity of the neural network. A neural network generally consists of input layers, convolution layer, pooling layers, fully connected layer, and output layers.

### 2.6. Segmentation model assessment

Intersection over Union (IoU) is a metric used to measure the degree of overlap between two sets. It is commonly employed to assess the the similarity between predicted and truth results in tasks such as image segmentation. IoU calculates the relative area rather than the absolute area, meaning that even if the absolute size of the predicted results and the true results differ, IoU can still provide a meaningful similarity measure. The definition of IoU is as follows:

$$\text{IoU} = \frac{(A) \cap (B)}{(A) \cup (B)} \tag{9}$$

Where (A) is the area of the predicted result and (B) is the area of the true result. The numerator is the intersection area of the two regions and the denominator is the union area of the two regions. The value of IoU ranges from 0 to 1, and the closer it is to 1 the higher the degree of overlap between the predicted outcome and the true result.

### 2.7. Classification model assessment

### 2.7.1. Confusion matrix

A confusion matrix is a popular model evaluation metric. The horizontal axis in the confusion matrix chart represents the types of predicted results for the samples, while the vertical axis represents the types of labels for the samples. Accuracy, specificity, and sensitivity closer to 1 indicate better diagnostic results of the constructed model. TPR represents sensitivity, FPR represents specificity, TP represents the number of positive samples in the validation set that the model correctly classified, FN represents the number of misclassified positive samples, FP represents the number of misclassified negative samples in the validation set, and TN represents the number of negative samples in the validation set.

$$Accuracy = \frac{TP+TN}{TP+TN+FP+FN} \tag{7}$$

$$Recall = \frac{TP}{TP+FN} \tag{8}$$

$$Precision = \frac{TP}{TP+FP} \tag{9}$$

$$\text{F1 Score} = \frac{2*\text{Precision}*\text{Recall}}{\text{Precision}+\text{Recall}} \tag{10}$$

$$\text{FPR} = \frac{\text{TP}}{\text{TN}+\text{FP}} \tag{11}$$

**2.7.2. ROC curve and AUC**

Receiver Operating Characteristic (ROC) curve and Area Under the Curve (AUC) are important tools for evaluating the performance of classification models. The ROC curve displays the trade-off between True Positive Rate (TPR) and False Positive Rate (FPR) at different classification thresholds, helping us understand the model's performance at various thresholds. AUC is the area under the ROC curve, usually ranging between 0.5 and 1. A higher AUC value indicates better model performance, while a value closer to 0.5 indicates poorer performance. It is a concise yet powerful metric for comparing the performance of different models, with a larger AUC typically indicating that the model can more accurately distinguish between positives and negatives.

## 3. Result

**3.1. Comparison of Segmentation Model**

After creating a standard UNet model and a specialized MultiUNet model for segmenting white matter in thick-slice MRI images, we fed preprocessed samples into the UNet network for training to automatically segment the samples. During testing, UNet showed an Intersection over Union (IoU) of 0.88, while MultiUNet achieved 0.92. Fig.5 shows the segmentation result of MultiUNet. It can be seen that the segmented white matter is closer to the ground truth. This result highlighted the superior performance of the MultiUNet network in segmenting white matter in thick-slice MRI images.

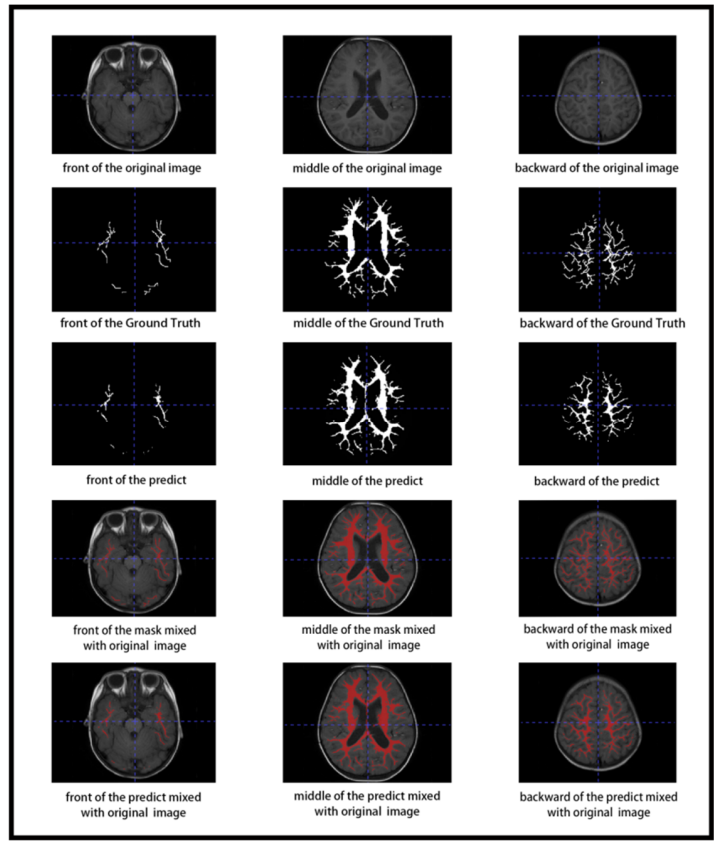

Fig.5.Segmentation result of MultiUNet

### 3.2. Comparison of Classification Model

Five typical machine learning models, including SVM, RF, KNN, LG CNN, were employed in this study to train the features and samples for comparison. Evaluating multiple models provided a more comprehensive understanding of their performance in ASD diagnosis, exploring the correlation between white matter lesions and ASD.

Table.1. The Performance of Each Classification Model

| Model | Input | AUC | Accuracy | F1-score | Specificity | Recall |
|---|---|---|---|---|---|---|
| SVM | White Matter Features | 0.89 | 0.8947 | 0.8889 | 0.8889 | 0.8889 |
| RF | White Matter Features | 0.85 | 0.8421 | 0.8125 | 0.7647 | 0.8667 |
| KNN | White Matter Features | 0.86 | 0.8684 | 0.8275 | 0.8571 | 0.8000 |
| LR | White Matter Features | 0.82 | 0.8157 | 0.8372 | 0.7200 | 1.0000 |
| CNN | Overall Image | 0.79 | 0.7631 | 0.7272 | 1.0000 | 0.5714 |
| CNN | Segmented White Matter | 0.84 | 0.8684 | 0.8148 | 0.9166 | 0.7333 |

Table.1 presents the metrics of the various models in terms of classification performance. These metrics include AUC, Accuracy, F1-score, Specificity, and Recall, etc. It can be observed that SVM exhibits the best overall performance, with the highest predictive set AUC (0.89), Accuracy (0.8947), and F1-score (0.8889). For the CNN, the accuracy of inputting segmented white matter reaches 0.8684, which has a higher accuracy compared to the whole image without segmentation (0.7631), highlighting the association between white matter regions and autism.

Fig.6 and Fig.7 show the confusion matrix and ROC curves of the six classification models, which visualize the classification effect of different models on different types of samples. LR does not classify normal people well, while CNN does not classify autistic people as well. The rest of the models are less affected by the sample imbalance problem and have superior classification performance for both positive and negative samples.

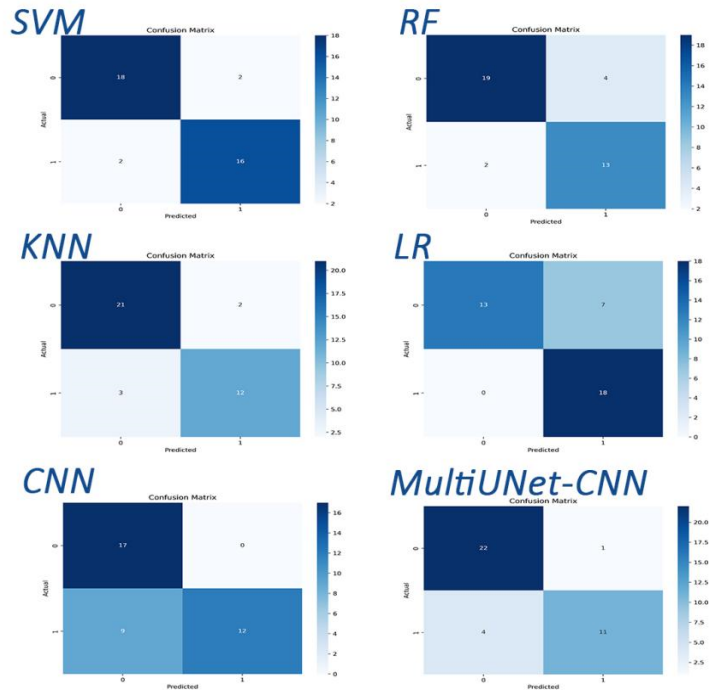

Fig.6. Confusion matrix

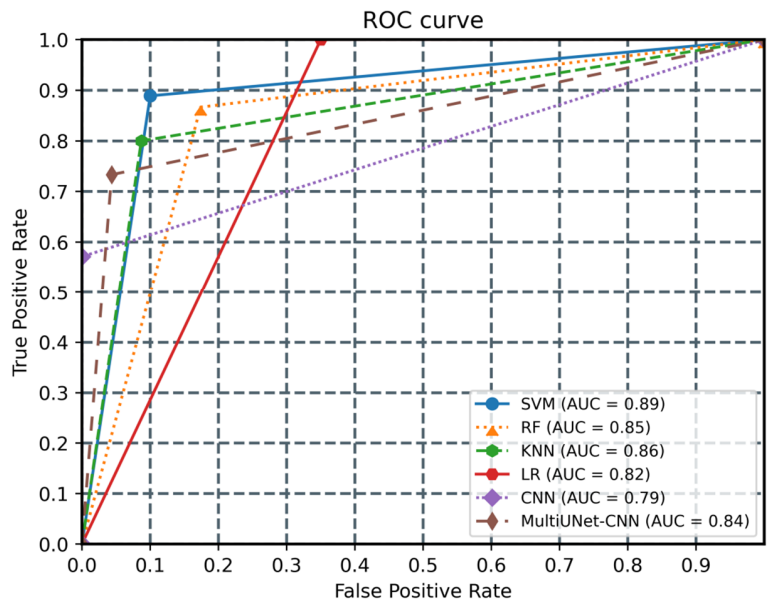

Fig.7. ROC curve and AUC

# 4. Discussion

Currently, a significant challenge in diagnosing ASD is the reliance on subjective behavioral observations and scale scores. To address this issue, more and more studies shift towards using neuroimaging to introduce a more objective and measurable aspect to the diagnostic process. However, most studies utilize overall brain MRI images for diagnosis which cannot fully utilize the association of specific brain regions with autism. Segregation of specific brain structures for classification can provide more precise and detailed diagnosis, as well as probe structural brain lesions in autism, providing deeper insights into the pathomechanisms of ASD. By training a segmentation model for automated white matter segmentation and utilizing its features to diagnose ASD , we were able to explore the correlation between white matter and ASD.

The study used thick-layer MRI images, but encountered segmentation undesirable segmentation results due to the large sampling interval and morphological differences between layers. To address this, a MultiUNet model was developed, consisting of three UNet models for segmenting different portions of the MRI images. This model design improved segmentation performance by efficiently capturing features in each layer. The MultiUNet model is expected to be applicable for segmenting 3D medical images and outperform traditional UNet models in tasks with large inter-layer intervals.

After segmenting the white matter, we utilized the Pyradiomics toolkit to extract MRI image features. These features were then fed into various machine learning models, including SVM, RF, LG, and KNN, resulting in an average accuracy of over 80% for ten-fold cross-validation. We also input the segmented brain white matter images into CNN, and the prediction accuracy also reached 86.84%. The SVM had the highest prediction accuracy of 89.47%. The accuracy achieved by our model demonstrates the potential of machine learning in facilitating the diagnosis of ASD and validates the correlation between white matter lesions and autism. However, we must also recognize the limitations of the study. Due to the small size of our dataset, the bias present in the dataset and the generalizability of the model across different populations remains to be improved. In addition, due to the dynamic nature of ASD, longitudinal studies are necessary to explore the effects on ASD during white matter development.

# 5. Conclusion

There have been many studies utilizing brain MRI images and machine learning techniques to achieve an objective diagnosis of ASD. However, most studies utilize overall brain MRI images for diagnosis, and the potential correlation between white matter and ASD has not been fully explored. To investigate this link and enable early and objective ASD diagnosis, We constructed a computer-aided diagnostic algorithm for early ASD diagnosis based on brain MRI white matter and machine learning methods. The MultiUNet model was trained to create a network for white matter segmentation, white matter features were extracted using an imaging histology method, and various machine learning models were compared. Among these models,

SVM demonstrated the highest prediction accuracy at 89.47%. This research offers a thorough evaluation of different models for ASD diagnosis, introduces a computer-aided diagnostic algorithm for early ASD diagnosis based on MRI white matter regions, and confirms a connection between white matter lesions and autism.

## Acknowledgements

The authors would like to thank all the reviewers who participated in the review. The authors would like to express their gratitude to the National Natural Science Foundation of China, Key Research and Development Program of Shandong Province, Department of Science and Technology of Shandong Province for funding and supporting this research. Special thanks to the dedicated medical professionals who contributed to the manual segmentation of MRI images, providing crucial Ground Truth data for the model training.

## Funding

This study has received funding by the National Natural Science Foundation of China(22176115), Key Technology Research and Development Program of Shandong Province(Major Scientific and Technological Innovation Project) (2021CXGC010506), Natural Science Foundation of Shandong Province (ZR2021MH208).

## Conflicts of interest

None of the authors have a conflict of interest to declare.

## Ethics approval

This study was conducted following the ethical guidelines outlined by the Institutional Review Board of Qilu Hospital, Shandong University.

## Consent to participate

Informed consent was obtained from all participants, and their privacy and confidentiality were strictly maintained throughout the research process.The study adheres to the principles of the Declaration of Helsinki.

## Consent for publication

Written informed consent for publication was obtained from all participants.

## Availability of data and code

All data, models, or code generated or used during the study are available from the corresponding author by request.